# Design of a fault free Inverter Circuit using Minimum number of cells & related Kink Energy Calculation in Quantum dot Cellular Automata

Ratna Chakrabarty
Department of Electronics & Communication
Institute of Engineering & Management
Kolkata, India
ratna_chakrabarty@yahoo.co.in

Angshuman Khan
Department of Electronics & Communication
Institute of Engineering & Management
Kolkata, India
angshumankhan2910@gmail.com

*Abstract*—Quantum dot Cellular Automata (QCA) is the most promising nanotechnology in the field of microelectronics and VLSI systems. QCA-based circuits require less power with a high switching speed of operation compared to CMOS technology. QCA inverter is one of the basic building blocks of QCA circuit design. The conventional QCA inverter design requires many cells. In this paper, we design the QCA inverter circuit using a lesser number of cells. We showed the kink energy calculation for the QCA-implemented inverters as well as the polarization of the circuits.

## I. INTRODUCTION

Quantum dot Cellular Automata (QCA) is an emerging technology that offers a revolutionary approach to nano-level computing [1]. QCA is the identical, bi-stable units or cells coupled through each other by electromagnetic forces [2]. Each cell consists of four to six dots with two extra electrons, resulting in two ground state configurations where the electrons occupy diagonal sites within the cell due to Coulombic interaction [3]. In digital technology, the logic levels are encoded as voltages on capacitors, whereas the information in QCA devices is encoded with the positions of the electrons within the cells. As a result, there is no current flow, and thus, there is no or little self-heating problem in a QCA circuit.

In this paper, we design an inverter using three QCA cells, though the conventional QCA inverter requires more than three QCA cells. Designing an inverter using two cells is also possible, but polarization becomes less for that design. So, we chose three cells to get a better response. We manually insert cells at the QCA inverter's output section to make the implemented three-cell inverter circuit more fault-free and find their corresponding kink energy. A simulation tool is essential to perform a defect characterization of QCA devices and circuits and to study their effects at the logic level. Here, we are using the QCADesigner tool [13], and from the simulated output, we point out the change of polarization of the inverter circuits.

## II. QCA OVERVIEW

### A. QCA Basics

The basic unit of QCA-based nanotechnology is a QCA Cell. A QCA cell is constructed from four to six quantum dots. The cell is charged with two free electrons and can tunnel between adjacent dots within the cells but not between the cells [10]. The electrons tend to occupy the corner positions by Coulombic repulsion [5, 6]; thus, two equivalent polarization states exist, namely, P = +1 and P = -1, shown in Fig. 1, with four quantum dots. These two polarization states represent logic '1' and logic '0', respectively.

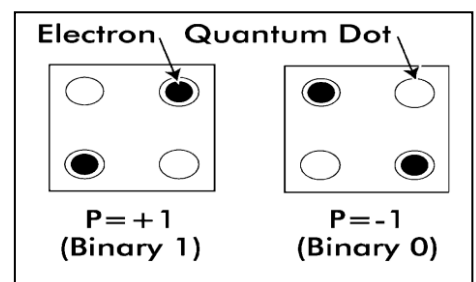

Fig. 1. QCA cells

### B. QCA Logic Devices

The fundamental QCA logic primitives are a QCA wire, QCA inverter, and QCA majority gate [4, 5], as described below.

In a QCA wire, the QCA cells are arranged along a straight line where the binary signal propagates from input to output through electrostatic interactions between cells. The propagation of information in the QCA wire is shown in Fig. 2.

A conventional QCA inverter circuit using eleven cells is shown in Fig. 3.

Another fundamental QCA logic circuit is the Majority gate, as shown in Fig. 4, which has three inputs and one output. The output cell will polarize to the majority polarization of the input cells. The center cell is the device cell. If A, B, and C are the inputs, the majority gate computes the function M = AB+BC+CA.

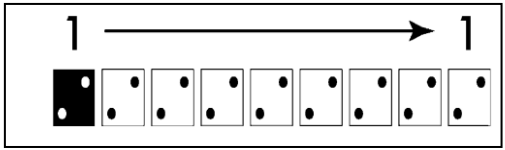

Fig. 2. QCA wire

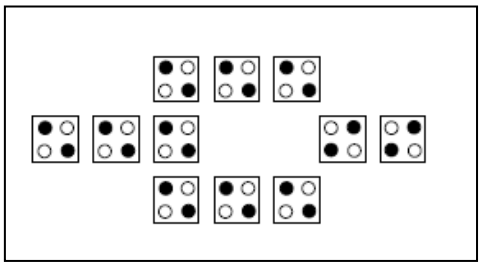

Fig. 3. Conventional QCA inverter

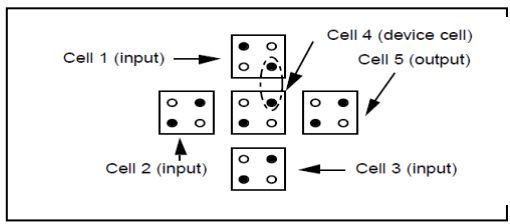

Fig. 4. QCA Majority Gate

### III. QCA CLOCKING MECHANISM

Two types of switching are possible in the operation of the QCA: abrupt switching and adiabatic switching. In abrupt switching, the changes in the input cell bring the system into an arbitrarily excited state with unstably high energy [7]. In adiabatic switching, the system is always kept in instantaneous ground state [8]. A clock signal is introduced to ensure it, and the signal either raises or lowers the tunneling barrier between the dots of a QCA cell. When the barrier is high, the cells are not allowed to change their state but are in a non-polarized state for the low barrier.

Figure 5 shows the four phases of adiabatic clocking: switch, hold, release, and relax. Each phase is separated by 90 degrees. The QCA circuit is partitioned into these four zones, and the same clock signal controls all cells in a zone [9]. In the switch phase, the actual computation occurs according to the input. In this phase, the inter dot potential barriers are first low, then barriers are raised, and the QCA cell becomes polarized to the state of its input driver. By the end of this clock phase, barriers are high enough, and there is no possibility of electron tunneling; hence, the cell state is fixed. In the second phase of clocking, barriers are still high and do not allow electron tunneling; hence, the output state is lowered, and electron positions are configured according to the original state. The cells are relaxed (unpolarized) in the last phase. So, the overall polarization of the QCA cell is determined in the switch and hold phase, where cells are in a polarized state depending upon the neighboring cell polarizations. In the release and relaxation phases, the states are unpolarized.

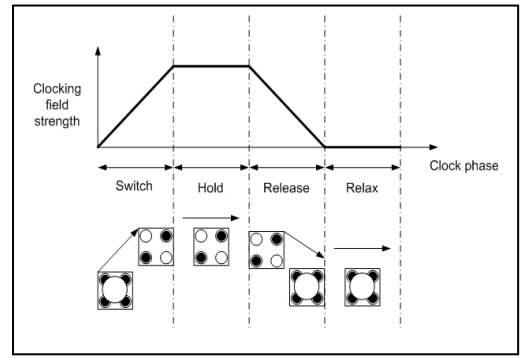

Fig. 5. Adiabatic clocking in QCA

### IV. KINK ENERGY

Due to the mutual repulsion of like charges, the two electrons in a four-dot QCA cell occupy diagonal positions. The electrostatic interaction between two QCA cells is given by

$$E = \frac{1}{4\pi\epsilon_0\epsilon_r} \cdot \frac{Q_1 Q_2}{r} = k \cdot \frac{Q_1 Q_2}{r}$$

The value of 'k' is $9 \times 10^{-9}$. As Q1 and Q2 are electronic charges. So, E is defined as

$$E = \frac{23.04 \times 10^{-29}}{r} J$$

This interaction determines the kink energy between two cells.

$$E_{kink} = E_{opp.polarization} - E_{same\ polarization}$$

Kink energy between two cells depends on the dimension of the QCA cell and the spacing between the cells, but it does not depend upon the temperature [10, 11].

### V. QCA INVERTER USING A MINIMUM NUMBER OF CELLS AND RELATED KINK ENERGY

The inverter can be constructed in QCA by only three cells, as shown in Fig. 6 (a). The symmetric design ensures exact symmetry between the inversion of a one and a zero. The signal that came from the leftmost is inverted at the point of convergence [12]. Fig 6(b) shows the inverter circuit using the QCADesigner tool, where 'a' is the input and 'b' is the output of the inverter. The corresponding output is shown in Fig. 6(c). The polarization from the simulation is found to be $\pm 9.50e^{-001}$, which tells that the inverter in Fig. 6 is faulty because the fault-free circuit has polarization near about $\pm 9.95e^{-001}$.

The calculated kink energy for the inverter shown in Fig. 6(a) is $6.838 \times 10^{-20}$ J.

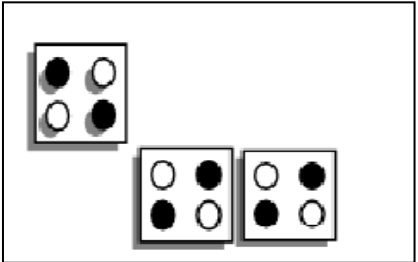

Fig. 6(a) QCA inverter using three cells

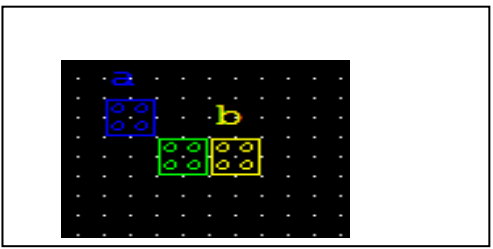

Fig. 6(b). Inverter implementation in QCADesigner tool

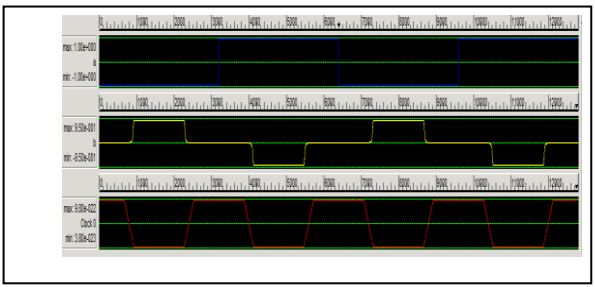

Fig. 6(c). Simulation result of the inverter is shown in Fig. 6(a)

## VI. ADDITION OF AN EXTRA CELL IN THE THREE CELLS QCA INVERTER

To minimize the fault of the inverter shown in Fig. 6(a), one extra cell is added to the output cell to make the inverter fault-free, as shown in Fig. 7(a). From the simulated output, as in Fig. 7(b), it is seen that the polarization becomes $\pm 9.86e^{-001}$, which proves that the inverter is almost fault-free.

The Kink energy for the inverter circuit shown in Fig. 7(a) is $10.862 \times 10^{-20}$ J.

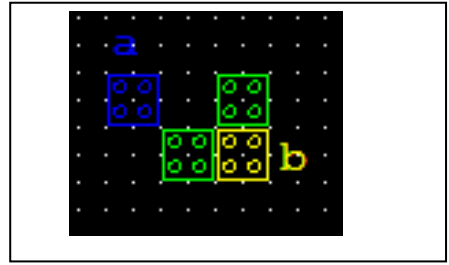

Fig. 7(a) Inverter using four cells

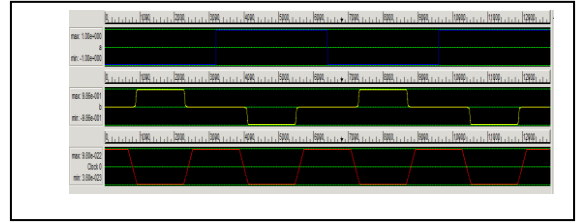

Fig. 7(b) Simulation result of inverter shown in Fig. 7(a)

## VII. ADDITION OF TWO EXTRA CELLS IN THE THREE CELLS QCA INVERTER

Fig. 8(a) shows the inverter after adding two extra cells with the output cell of the faulty inverter shown in Fig. 6(a). The polarization in the simulated output shown in Fig. 8(b) becomes $\pm 9.94e^{-001}$. The calculated kink energy for the inverter circuit shown in Fig. 8(a) is $14.986 \times 10^{-20}$ J.

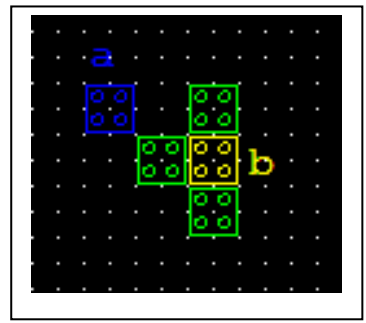

Fig. 8(a) Inverter using five cells

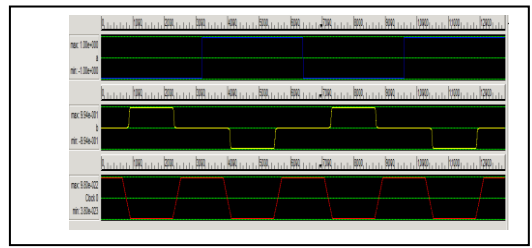

Fig. 8(b) Simulation result of inverter shown in Fig. 8(a)

## VIII. ADDITION OF THREE EXTRA CELLS IN THE THREE CELLS QCA INVERTER

In Fig.9 (a), three extra cells are added to the output cell of the faulty inverter, as shown in Fig. 6(a). It makes the circuit fault-free. The polarization in the simulated output shown in Fig.9 (b) is $\pm\ 9.94e^{-001}$, and the kink energy of the output is $17.328 \times 10^{-20}$ J.

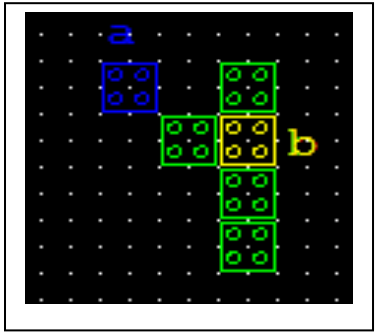

Fig. 9(a) Inverter using six cells

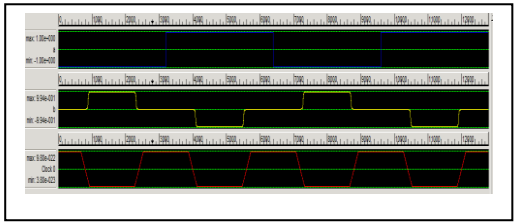

Fig. 9(b) Simulation result of inverter shown in figure 8(a)

The Kink energy & polarization of the above-discussed inverter circuits are shown in tabular form in TABLE I. Here, the kink energy and polarization increase with the number of cells in the initial inverter QCA circuit consisting of three cells. However, the polarization is not further increased by adding a greater number of cells after the last calculation. So, from this result, a QCA inverter can be made of five cells without an error.

The plot of Kink energy vs. number of cells and Maximum Polarization vs. number of cells are shown in Fig. 10 and Fig. 11, respectively.

TABLE I. KINK ENERGY CALCULATION

| Sl. No. | No. of cells | Kink energy(J) | Polarization |
|---|---|---|---|
| 1. | 3 | $6.838 \times 10^{-20}$ | $\pm 9.50e^{-001}$ |
| 2. | 4 | $10.862 \times 10^{-20}$ | $\pm 9.86e^{-001}$ |
| 3. | 5 | $14.986 \times 10^{-20}$ | $\pm 9.94e^{-001}$ |
| 4. | 6 | $17.328 \times 10^{-20}$ | $\pm 9.94e^{-001}$ |

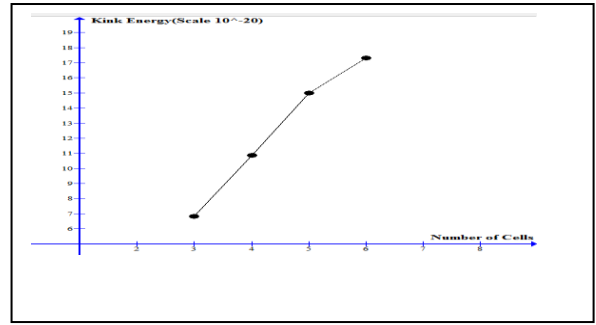

Fig. 10. Plot of Kink energy vs. number of cells

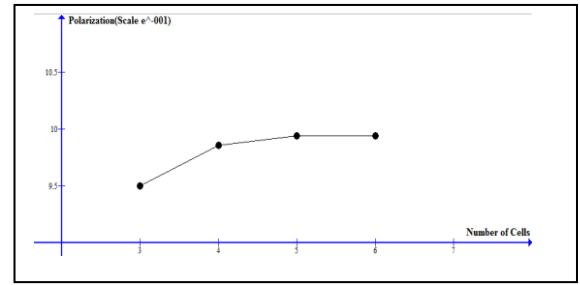

Fig. 11. Plot of Maximum Polarization vs. number of cells

## IX. CONCLUSION

From the above results, it is concluded that the QCA inverter can be constructed using only three cells, which require less area than the conventional inverters consisting of more cells. To make the inverter fault-free, it is required to add cells at the output section of the circuit. Though the cell number increases, it is still less than the conventional one. The area consumption of the circuit is also much less, which is helpful in designing the larger circuit. The Kink energy and the polarization of the inverter are increased with the increase of the number of additional cells, and it is observed that the polarization of the circuit is constant after five cells, so a QCA inverter can be made with five cells without any error.